\documentclass[journal,twoside,web]{ieeecolor}
\usepackage{jsen}
\usepackage{cite}
\usepackage{amsmath,amssymb,amsfonts}
\usepackage{array} 
\usepackage{algorithm}
\usepackage{algorithmicx}
\usepackage{graphicx}
\usepackage{textcomp}
\usepackage{wrapfig}
\usepackage{algpseudocode}
\usepackage{setspace}
\usepackage{subfigure}
\usepackage{caption}
\usepackage{xcolor}

\def\BibTeX{{\rm B\kern-.05em{\sc i\kern-.025em b}\kern-.08em
    T\kern-.1667em\lower.7ex\hbox{E}\kern-.125emX}}
\definecolor{abstractbg}{rgb}{0.89804,0.94510,0.83137}
\begin{document}
\title{{Tradeoff Between the Number of Transmitted Molecules and the BER Performance in Molecular Communication between  Bionanosensors}}
\author{Dongliang Jing, \IEEEmembership{Member, IEEE}, Linjuan Li, Lin Lin, \IEEEmembership{Senior Member,IEEE} and Andrew W. Eckford, \IEEEmembership{Senior Member,IEEE}
\thanks{This work was supported by China Postdoctoral Science Foundation under Grant 2023M732877, and in part by the Natural Science Basic Research Program
of Shaanxi under Program 2024JC-YBQN-0649.}
\thanks{Dongliang Jing is with the College of Mechanical and Electronic Engineering, Northwest A\&F University, Yangling, China, and the Key Laboratory of Agricultural Internet of Things, Ministry of Agriculture and Rural Affairs, Yangling, China (e-mail: dljing@nwafu.edu.cn).}
\thanks{Linjuan Li is with the College of Mechanical and Electronic Engineering, Northwest A\&F University, Yangling, China (e-mail: 2021012605@nwafu.edu.cn).}
\thanks{Lin Lin is with the College of Electronics and Information Engineering, Tongji University, Shanghai, China (e-mail: fxlinlin@tongji.edu.cn).}
\thanks{Andrew W. Eckford is with the Department of Electrical Engineering and Computer Science, York University, Toronto, Ontario, Canada (e-mail: aeckford@yorku.ca).}}

\IEEEtitleabstractindextext{%
\fcolorbox{abstractbg}{abstractbg}{%
\begin{minipage}{\textwidth}%
\begin{abstract}
In the domain of molecular communication (MC), information is conveyed through the characteristics of molecules transmitted between the transmitter and the receiver bionanosensors via propagation. The constrained size of the transmitter imposes limitations on its storage capacity, constraining the number of available molecules for transmission, with a resulting effect on communication reliability. This paper primarily focuses on achieving an equilibrium between the number of transmitted molecules and the bit error rate (BER) performance.
To this end, we first analyze the relationship between the number of transmitted molecules and the BER performance. Subsequently, a balancing function that considers both the number of transmitted molecules and the BER performance is introduced, taking into account the molecules' respective weights. Given the difference in magnitude between the number of transmitted molecules and the BER, these parameters are normalized to facilitate analysis. Subsequently, a Gradient Descent Algorithm is employed to determine the optimal number of transmitted molecules, aiming to achieve the optimal equilibrium in the analyzed MC system.
Theoretical and simulation results are provided, substantiating that the optimal outcome indeed establishes an ideal balance between the number of transmitted molecules and the BER. 
\end{abstract}

\begin{IEEEkeywords}
Molecular communication (MC), transmitted molecules, bit error rate (BER), balance function. 
\end{IEEEkeywords}
\end{minipage}}}

\maketitle

\section{Introduction}
\label{sec:introduction}
\IEEEPARstart{W}{ith} recent advancements in nanotechnology, biotechnology, and synthetic biology, the realization of bionanosensors, also known as nanomachines, has become feasible. However, due to their inherent size limitations, nanomachines are relegated to performing rudimentary functions, such as sensing, computation, and actuation. Hence, the need arises for these nanomachines to communicate, forming intricate nano-networks \cite{farsad2016comprehensive,saeed2023novel,jing2024energy}. Molecular communication (MC), drawing inspiration from the innate communication mechanisms found in living cells, represents a powerful tool for exploring information exchange among nanomachines. In MC, information transfer relies on the exchange of molecules between the transmitter and the receiver. The study referenced in \cite{10443866} delves into the role of microfluidic technology in advancing MC, emphasizing the evolution from theoretical models to practical applications. MC envisions a wide range of applications, including military, ubiquitous healthcare, entertainment, and many other fields \cite{yang2020comprehensive,khaloopour2021theoretical,zhang2024design}.

In MC, molecules such as DNA, calcium ions, and neurotransmitters are typically used to transfer information. In \cite{xiao2023really}, the authors examined the intricacies of commonly used information molecules and their fundamental physical characteristics. Information can be encoded in the concentration of molecules, the type of molecules \cite{kuran2011modulation}, the release time of molecules \cite{4298292}, and other methods. In \cite{li2019asymmetric}, the information symbols are encoded based on the timing of molecule release, with each information bit being conveyed by the emission of a single molecule. Once the information molecules are released by the transmitter, the process can be modeled as either the unbinding of the information molecules from the transmitter or the opening of a molecular gate to allow the molecules to diffuse away \cite{nakano2012molecular}. During the propagation of information molecules from the transmitter to the receiver, the molecules can move by Brownian motion without consuming energy in diffusion-based MC. In \cite{nakano2017molecular}, the dynamic properties of oscillating and propagating patterns in the concentration of information molecules were studied. At the receiver, considering the importance of measuring the concentration of molecules in MC, the authors in \cite{moore2014diffusion} examined the impact of multiple measurements of the concentration of molecules on MC performance. Moreover, \cite{li2019clock} introduces a clock-free asynchronous receiver design, a detector that recovers information symbols without measuring the arrival time of molecules.

In MC, the availability of molecular resources is limited by the finite number of stored molecules in the transmitter and the constrained production rate due to chemical reactions. Molecules can be harvested from the surrounding environment or synthesized by the transmitter itself \cite{bi2021survey}. To address these limitations, various strategies have been proposed and studied. To improve energy efficiency, \cite{deng2016enabling} proposed an information molecule synthesis process for MC with relay, where the relay generates emission molecules from absorbed molecules via chemical reactions.  In \cite{musa2020feedback}, a diffusion-based MC system powered by a nanoscale energy-harvesting mechanism was examined, along with a power control mechanism based on feedback control theory. The process of molecule harvesting in neurons, known as re-uptake, was investigated and modeled using stochastic chemical reactions in \cite{lotter2020synaptic}. Additionally, \cite{ahmadzadeh2022molecule} considered a transmitter equipped with molecule harvesting units on its surface, allowing signaling molecules that contact these units to be re-captured. This study also derived a closed-form expression for the harvesting impulse response. Reference \cite{huang2023analysis} explored a molecule harvesting transmitter model with heterogeneous receptors covering the transmitter's surface. These receptors absorb any molecules that hit them, further enhancing the efficiency of molecular resource utilization.

Furthermore, the molecule release rate by the transmitter has been investigated in various studies. In \cite{wen2024absorption}, the authors proposed an absorption shift keying scheme for MC, where molecules serve both as information carriers and energy providers through chemical reactions. This scheme involves harvesting unused molecules to enhance system performance and introduces a third switch-controllable molecule harvesting node, in addition to the conventional transmitter and receiver, in a point-to-point MC scenario. Inspired by natural communications among living cells, which have the ability to collect and store food or energy from the environment to produce information molecules, \cite{kumar2024game} explored a scenario with multiple transmit nodes interacting within an environment. Sharing the available common food molecular budget information is critical to the overall system performance. The authors employed game theory to investigate the effects of behavioral interactions among transmitter nanomachines.

In an MC system, a group of bio-nanomachines can act as receivers and chemically react to molecules propagating through the environment.  However, if the release rate is faster than the receiver's reaction time, it can lead to a degradation in efficiency. To address this, \cite{nakano2013transmission} studied optimal transmission rates that maximize both throughput and efficiency. Considering the limited molecule production and storage, \cite{khaloopour2018adaptive} proposed an adaptive pulse-width modulation scheme by varying the duration of molecule release. In \cite{cheng2023channel}, Cheng et al. minimized the average bit error rate (BER) of a multiple-input multiple-output mobile molecular communication system by optimizing the number of released molecules by each transmitter. \cite{panahi2024energy} focused on optimizing the energy efficiency of a molecular data-collection nanonetwork with energy-constrained nanosensors, taking into account constraints on molecular concentration, data rate, and available molecular resources. To improve the performance of multi-hop mobile MC with drift, \cite{cheng2022joint} adopted an amplify-and-forward relay strategy, using multiple and single molecule types in each hop to transmit information. Considering the application of diffusion-based nano-sensor networks for data gathering in in-body medical systems and the limited capabilities of the nano-sensors, \cite{shitiri2021tdma} studied a lightweight time-division multiple access (TDMA) based data gathering multiple access control protocol.

In \cite{li2019novel}, a time-based modulation scheme is proposed, which can also be applied to scenarios involving the transmission of multiple information molecules. The study demonstrates that the BER performance can be significantly enhanced by increasing the number of released information molecules per bit in the considered MC system. Unlike \cite{chouhan2020gradient}, which considered a cooperative communication system where the optimal number of molecules transmitted from the source and cooperative nanomachines was obtained to minimize the error probability with a fixed decision threshold and a constant molecular budget, this paper addresses the challenges of limited molecular resources and varying communication reliability requirements in different MC systems. We first analyze the relationship between the BER performance and the number of transmitted molecules. To achieve a trade-off between the number of transmitted molecules and BER performance, we establish a balance function that jointly considers BER performance and the number of transmitted molecules with different weights. In MC applications requiring high communication reliability, a larger number of molecules are transmitted. Conversely, in scenarios where the number of available molecules is strictly limited and high communication reliability is not essential, a smaller number of molecules are transmitted. Such as in drug delivery systems, the ability to precisely control the number of molecules released is vital for accurate targeting within the body. This is especially important in treatments like cancer therapy, where high accuracy and low BER are critical for ensuring that the therapeutic agents reach the target site with minimal errors, thus maximizing treatment efficacy. Conversely, in situations where delivery precision is less critical or molecular resources are limited—such as in long-term drug administration—releasing fewer molecules may be more practical. This approach conserves resources while still maintaining an acceptable level of performance, effectively balancing therapeutic impact with resource management.

The main contributions of this paper are summarized as follows:
\begin{itemize}
\item A detailed analysis of the relationship between BER performance and the number of transmitted molecules in diffusion-based MC systems is provided, offering valuable insights into the optimization of molecular communication.
\item A balance function is developed that integrates considerations of BER performance and the number of transmitted molecules by normalizing these two parameters to address the constraints of limited molecular resources in the transmitter.
\item A Gradient Descent Algorithm is conducted based on the established balance function to identify the optimal number of transmitted molecules, ensuring that communication reliability requirements are met efficiently.
\end{itemize}

The remainder of this paper is organized as follows. In Section II, we introduce the system model of the considered MC. In Section III, the tradeoff between the number of transmitted molecules and BER performance is analyzed. Numerical and simulation results are presented in Section IV. Finally, Section V concludes this paper.

\section{System Model}
\begin{figure}[!t]
   \centering
   \includegraphics[width=0.5\textwidth]{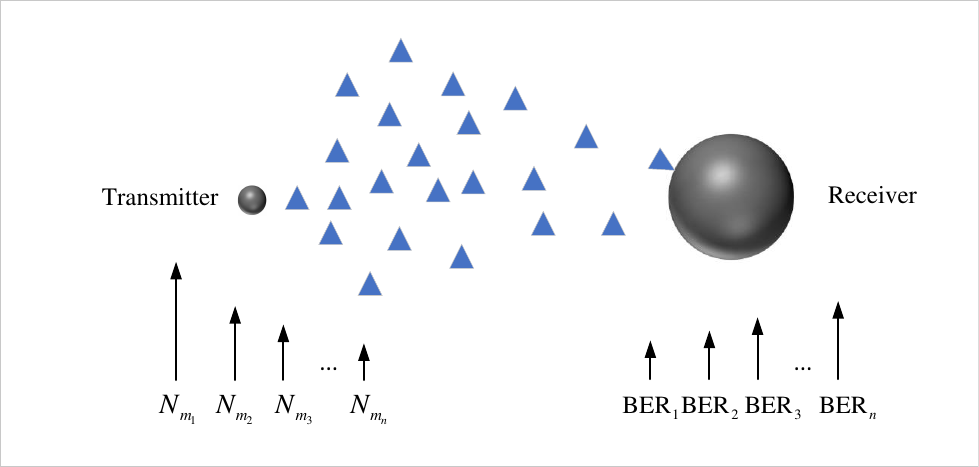}\\
   \caption{The system model of the considered MC system, where $N_m$ denotes the number of transmitted molecules, and BER represents the corresponding bit error rate, shows that different numbers of transmitted molecules result in varying BER performance.}\label{system_model} 
\end{figure}

In this paper, we consider a 3D diffusive MC system with a point transmitter and a spherical passive receiver. Information is encoded in the concentration of molecules, which are then propagated to the receiver through a diffusion channel. The information molecules are measured within the detection space based on the time-varying molecular concentration at the receiver. The detection space of a spherical passive receiver refers to the spherical region surrounding the receiver, where molecules are observed and detected. This model is commonly used and widely studied in MC \cite{pierobon2011diffusion, kuscu2019transmitter, trinh2021molecular}. The system model is shown in Fig. \ref{system_model}.

At the transmitter, On-Off keying (OOK) modulation is employed to encode the information, which means to transmit bit 1, $N_m$ molecules are released, while to transmit bit 0, no molecule is released, the transmitted bit $b_{tx}$ can be expressed as
\begin{equation}
b_{tx}
= \left\{ {\begin{array}{*{20}{c}}
{1,}&{N_{tx}=N_m}\\
{0,}&{N_{tx}=0}
\end{array}\begin{array}{*{20}{c}}.\\
\end{array}} \right.
\end{equation}

Therefore, the transmitted signal of the released molecules during the $k$th bit interval can be expressed as
\begin{align}
{s_{tx,k}}\left( t \right) = {N_m}\delta \left( {t - kT_b} \right),
\end{align}
where $\delta()$ is the Dirac delta function and  $T_b$ is the bit interval.

Assuming the distance between the transmitter and the point receiver is $d$, the radius of the receiver is $r$, and the receiver is able to measure the concentration within the spherical space defined by this radius. Additionally, assuming the transmitter and the receiver are perfectly synchronized. Therefore, after a pulse of molecules is released at time $t=0$, then, based on Fick's second law, at time $t$, the molecular impulse response observed by the receiver $h(t)$ can be expressed as
\begin{align}
h\left( t \right) = \frac{1}{{{{\left( {4\pi Dt} \right)}^{{3 \mathord{\left/
 {\vphantom {3 2}} \right.
 \kern-\nulldelimiterspace} 2}}}}}\exp \left( { - \frac{{{d^2}}}{{4D_mt}}} \right),
\end{align}

After a single impulse of molecules is released at the beginning of the bit interval, the concentration at the receiver increases to its peak and then decreases with a long tail. The time corresponding to the peak concentration can be determined by finding the time at which the derivative of $h(t)$ with respect to $t$ is 0 and can be expressed as
\begin{align}
\frac{{\partial h\left( t \right)}}{{\partial t}} = 0.
\end{align}
Using this expression, the peak time can be achieved at $t_p=d^2/6D_m$, while the corresponding peak concentration $c_max$ is 
\begin{align}
{c_{\max }} = \frac{{{N_m}}}{{{d^3}}}{\left( {\frac{3}{{2\pi e}}} \right)^{{3 \mathord{\left/
 {\vphantom {3 2}} \right.
 \kern-\nulldelimiterspace} 2}}}.
\end{align}
Therefore, the bit interval $T_b$ should be larger than $t_{\rm{peak}}$ to ensure that the peak concentration occurs within the bit interval. The concentration of the received molecules without noise can be expressed as
\begin{align}
\begin{split}
y\left( t \right) &= s\left( t \right) * h\left( t \right)\\
 &= {N_m}h\left( {T_b} \right).    
\end{split}
\end{align}

Considering the counting noise $n_k$, which arises due to the randomness of molecular propagation and the discrete nature of the molecular diffusion process, and follows a normal distribution  $n_k\sim N(0,\sigma_{n}^2)$. The variance $\sigma_{n}^2$ can be expressed as
\begin{align}
\sigma_n^2=\frac{y(t)}{V_{\rm{rx}}}.  
\end{align}

Based on the above analysis, in MC, the Signal-to-Noise Ratio (SNR) can be expressed as \cite{wang2015distance}
\begin{align}
\label{SNR}
\begin{split}
{\rm{SNR}}&=\frac{y^2(t)}{\sigma_n^2}\\
&={N_m}{V_{\rm{rx}}}h\left( {T_b} \right). 
\end{split}
\end{align}

As depicted in (\ref{SNR}), the SNR increases with the number of transmitted molecules $N_m$. However, increasing the number of transmitted molecules $N_m$ requires more energy for the transmitter to generate these information molecules and is constrained by the limited supply of molecules in the transmitter. Through a higher SNR, the receiver's BER performance is enhanced.

In diffusion-based MC, after the molecules are released by the transmitter, they propagate to the receiver via Brownian motion, following different trajectories. The remaining molecules in the channel affect the concentration at the receiver of the newly transmitted molecules, leading to inter-symbol interference (ISI). Considering the counting noise, the molecule concentration at the receiver at time $t$, $\left(k-1\right)T_b \le t <  {k } T_b$ by transmitting a bit sequences $b_j = \{0,1\}$ can be expressed as
\begin{align}
y_n\left( t \right) = \sum\limits_{i = 1}^k {{b_i}{N_m}h\left( {t - (i-1)T_b} \right)} + n(t),
\end{align}

Assuming the receiver samples the received molecules at the peak concentration $t_{p}$ during $[(k-1)T_b, kT_b]$, then the received signal by the receiver during the $k$th bit interval can be expressed as
\begin{align}
\label{yk_isi}
y_k = \sum\limits_{i = 1}^k {{b_i}{N_m}h(\left( {k-i} \right)T_b+t_p)} + n(k),
\end{align}
In diffusion-based MC, it is often observed that the majority of ISI is contained within a finite number of time slots, beyond which the influence of previous ISI can be neglected \cite{sabu2019analysis}. Hence, by considering a finite length of ISI, (\ref{yk_isi}) can be expressed as
\begin{align}
y_k = \sum\limits_{i=k-L}^k {{b_i}{N_m}h(\left( {k-i} \right)T_b+t_p)} + n_k,
\end{align}

At the receiver, the received signal is decoded by comparing the received concentration to a preset threshold $C_{\rm{Thr}}$, and $C_{\rm{Thr}}$ should be smaller than $c_{\rm{max}}$. If the concentration of received molecules exceeds the threshold $C_{\rm{Thr}}$, the received signal is decoded as bit 1; otherwise, it is decoded as bit 0. This decision criterion can be expressed as
\begin{align}
\begin{split}
{b_{rx}} = \left\{ {\begin{array}{*{20}{c}}
{1,}&{{y_k} \ge {C_{\rm{Thr}}}}\\
{0,}&{\rm{otherwise}}
\end{array}} \right.    
\end{split}.
\end{align}

The received signal can be divided into three parts: the molecules released in the current bit interval and received in the current bit interval, expressed as $y_{k,c} = {b_k}{N_m}h(t_p)$; the molecules released from previous bit intervals but received in the current bit interval (within the memory length), expressed as $y_{k,\rm{ISI}} = \sum\limits_{i=k-L}^{k-1} {{b_i}{N_m}h(\left( {k-i} \right)T_b+t_p)}$; and the counting noise $n_k$. As the number of transmitted molecules is large in MC, then, at the receiver, the concentrations of $y_{k,c}$ and $y_{k,\rm{ISI}}$ can both be approximated by the Gaussian distributions. Therefore, the total concentration at the receiver can also be approximated by the Gaussian distribution. Based on the above analysis, and considering the fact that the number of transmitted molecules $N_m$ is large, during the $k$th bit interval, the received molecules $y_{k,c}$ follow a normal distribution and can be expressed as
\begin{align}
\label{y_c_distribution}
y_{k,c} \sim \mathcal{N}\left({b_k{N_m}{h_1},{\frac{1}{V_R}}b_k{N_m}{h_1}(1 - {h_1})} \right), 
\end{align}
where $h_1=h(t_p)$, and $t_p$ is the maximum molecule concentration measured by the receiver after the molecules released by the transmitter at time $t=0$. Similarly, the molecules transmitted at the beginning of $i$th bit interval and received during the $k$th bit interval ($i<k$) $y_{k,{\rm{ISI}}}$, namely ISI can be expressed as:
\begin{equation}
\label{y_isi_distribution}
y_{k,i,{\rm{ISI}}} \sim \mathcal{N}\left( {{b_ i{N_m}}{h_{k-i}},{\frac{1}{V_R}}b_i{N_m}{h_{k-i}}(1 - {h_{k-i}})} \right) .
\end{equation}

\section{Tradeoff between the number of transmitted molecules and BER performance}

In this section, we address the challenge posed by the decrease in SNR as the number of transmitted molecules increases, which consequently enhances the BER performance. Nevertheless, it's imperative to acknowledge the limited molecular resources available in MC. Hence, we introduce a balance function that strategically considers both the quantity of transmitted molecules and BER performance, aiming to strike an optimal equilibrium between these two factors. 
\subsection{BER analysis}
In this section, considering the OOK scheme is utilized within the diffusion-based MC. At the receiver, errors arise due to factors such as ISI and counting noise. Referencing \cite{8412141}, the BER of OOK modulation in MC can be expressed as follows
\begin{align}
\begin{split}
{P_e} &= \frac{1}{2}P\left( {{{\hat b}_k} = 0|{b_k} = 1} \right) + \frac{1}{2}P\left( {{{\hat b}_k} = 1|{b_k} = 0} \right)\\
 &= \frac{1}{2}\left( {{P_M} + {P_{FA}}} \right)
\end{split}
\end{align}
where $P_M$ represents the probability of a miss, indicating that the transmitted bit $b_{tx}=1$ but the detected bit is $b_{rx}=0$, while $P_{FA}$ denotes the false alarm probability, indicating that the transmitted bit $b_{tx}=0$ but the detected bit is $b_{rx}=1$. The expressions for $P_M$ and $P_{FA}$ can be expressed as 

\begin{align}
\begin{split}
\label{P_M}
{P_M} &= 1 - \frac{1}{{{2^L}}}\sum\limits_{{b_L} \in \{0,1\}} {Q\left( {\frac{{{C_{\rm{Thr}}} - {\mu _{{Z_1}}}}}{{{\sigma _{{Z_1}}}}}} \right)} ,
\end{split}
\end{align}
and
\begin{align}
\begin{split}
\label{P_FA}
{P_{FA}} &= \frac{1}{{{2^L}}}\sum\limits_{{b_L} \in \{0,1\}} {Q\left( {\frac{{{C_{\rm{Thr}}} - {\mu _{{Z_0}}}}}{{{\sigma _{{Z_0}}}}}} \right)}.
\end{split}
\end{align}
respectively. Where $Q(x)$ is the tail distribution function of the standard normal distribution and can be expressed as 
\begin{align}
Q\left( x \right) = \frac{1}{{\sqrt {2\pi } }}\int_x^\infty  {\exp \left( { - \frac{{{x^2}}}{2}} \right)} dx.   
\end{align}
Moreover, in (\ref{P_M})-(\ref{P_FA}), ${\mu _{{Z_1}}}$ and $\sigma _{{Z_1}}^2$ are mean and variance of the received molecules under the binary hypothesis testing problem of ${\mathcal{H}_1}$; ${\mu _{{Z_0}}}$ and $\sigma _{{Z_0}}^2$ are mean and variance of the received molecules under the binary hypothesis testing problem of ${\mathcal{H}_0}$, and can be expressed as \cite{8412141}
\begin{align}
\begin{split}
{\mu _{{Z_1}}} &= {N_m}{h_1} + \sum\limits_{i = k-L}^k {{b_{i}}{N_m}h(\left( {k-i} \right)T_b+t_p)}, \\
\end{split}
\end{align}
\begin{align}
\begin{split}
\sigma _{{Z_1}}^2 &= \frac{{{b_1{N_m}h_1}}}{{{V_R}}} + \frac{1}{{{V_R}}}\sum\limits_{i = k-L}^k {{b_{i}}{N_m}h(\left( {k-i} \right)T_b+t_p)}, \\
\end{split}
\end{align}
\begin{align}
\begin{split}
{\mu _{{Z_0}}} &= \sum\limits_{i = k-L}^k {{b_{i}}{N_m}h(\left( {k-i} \right)T_b+t_p)}, \\
\end{split}
\end{align}
and
\begin{align}
\begin{split}
\sigma _{{Z_0}}^2 &= \frac{1}{{{V_R}}}\sum\limits_{i = k-L}^k {{b_{i}}{N_m}h(\left( {k-i} \right)T_b+t_p)}, 
\end{split}
\end{align}
respectively. And the receiver detects the received samples at time $t=kT_b+t_p$.

\subsection{Tradeoff between the number of transmitted molecules and BER performance}

%
%
In this subsection, we confront the challenge of an increasing SNR as the number of transmitted molecules increases, which results in a lower BER. However, this presents a dilemma as it requires transmitting more molecules, a resource that is 
scarce in the MC system. To address this issue, we propose a mathematical model that jointly considers the number of transmitted molecules and BER performance. Additionally, we investigate an optimization algorithm aimed at achieving a better balance between the number of transmitted molecules ($N_m$) and the BER ($P_e$).

Given the substantial difference in magnitude between $N_m$ and $P_e$, we first normalize these parameters to facilitate analysis. Considering a series number of transmitted molecules $[N_{m_1}, N_{m_2}, \cdots, N_{m_n}]$ and the corresponding BER values $[P_{e_1}, P_{e_2}, \cdots, P_{e_n}]$, the normalized values of transmitted molecules $ \hat{N}_m$ and the corresponding BER values $\hat{P}_e$, can be expressed as
\begin{align}
\hat{N}_m=\frac{{{N_m}\left( i \right) - \min \left( {{N_m}} \right)}}{{\max \left( {{N_m}} \right) - \min \left( {{N_m}} \right)}},
\end{align}

\begin{align}
\hat{P}_e=\frac{{{P_e}\left( i \right) - \min \left( {{P_e}} \right)}}{{\max \left( {{P_e}} \right) - \min \left( {{P_e}} \right)}}.
\end{align}

Then, the balance function $f_{balance}$ can be defined as
\begin{align}
\label{balance_function}
\begin{split}
f_{balance} =& {w_{N_m}} \times \hat{N}_m + {w_{P_e}} \times \hat{P}_e,     
\end{split}   
\end{align}
where $w_{N_m}$ and $w_{P_e}$ are the weights of $\hat{N}_m$ and $\hat{N}_m$, respectively. As shown in (\ref{balance_function}), the balance function jointly considers the normalized number of transmitted molecules and the normalized BER. This approach effectively balances the trade-off between the quantity of transmitted molecules and BER performance, allowing for optimal resource allocation in MC systems.

Then, the gradient descent algorithm is employed to optimize the number of transmitted molecules and achieve a balance between $N_m$ and $P_e$. The gradient descent algorithm is a widely used optimization algorithm that iteratively adjusts the parameters of a model to minimize a given objective function \cite{liu2019efficient,chouhan2020gradient,chen2021robust}. Specifically, the technique calculates gradients that reflect how changes in $N_m$ and $P_e$ impact the objective function, allowing for targeted adjustments. By updating the parameters based on these gradients, the algorithm efficiently searches for the optimal solution within the parameter space. The process continues until convergence, effectively balancing $N_m$ and $P_e$ and ultimately enhancing the performance of the proposed model.

To efficiently update the parameters of the balance function during the gradient descent optimization process, we first compute the partial derivative of the balance function with respect to $N_m$ as
\begin{align}
\begin{split}
\frac{{\partial {f_{balance}}}}{{\partial {N_m}}} =& {w_{{N_m}}}\frac{{\partial \hat{N}_m}}{{\partial {N_m}}} + {w_{{P_e}}}\frac{{\partial \hat{P}_e}}{{\partial {N_m}}},   
\end{split}
\end{align}
where
\begin{align}
\frac{{\partial \hat{N}_m}}{{\partial {N_m}}} = \frac{1}{{{N_{{m_{\max }}}} - {N_{{m_{\min }}}}}},
\end{align}
and
\begin{align}
\frac{{\partial \hat{P}_e}}{{\partial {N_m}}} = \frac{1}{{\max \left( {{P_e}} \right) - \min \left( {{P_e}} \right)}}\frac{{\partial {P_e}}}{{\partial {N_m}}}.
\end{align}
The derivative $\frac{{\partial {P_e}}}{{\partial {N_m}}}$ is given by
\begin{align}
\label{partial_{P_e}}
\frac{{\partial {P_e}}}{{\partial {N_m}}} = \frac{1}{2}\left( {\frac{{\partial {P_M}}}{{\partial {N_m}}} + \frac{{\partial {P_{FA}}}}{{\partial {N_m}}}} \right)
\end{align}
The detailed derivation of $\frac{{\partial {P_e}}}{{\partial {N_m}}}$ and $\frac{{\partial {P_M}}}{{\partial {N_m}}}$ is shown in the appendix.

Then, the Gradient Descent method is introduced to optimize the balance between the number of transmitted molecules $N_m$ and the bit error rate $P_e$ in the diffusion-based MC system. This algorithm is designed to minimize a given balance function by iteratively adjusting the values of $N_m$ and $P_e$. The pseudocode for the gradient descent method is provided in Algorithm I.

\begin{algorithm}
\setstretch{1.5}
\caption{Pseudocode of the Gradient Descent Algorithm}
\label{gradient_descent}
\begin{algorithmic}[1]
\Function{Gradient Descent}{$f_{balance}$, $w_{N_m}$, $w_{P_e}$, $N_m$, $P_e$, \text{learning\_rate}, \text{max\_iterations}, \text{tolerance}}
    \State Initialize: Initial the number of transmitted molecules $N_m$
    \State $iteration \gets 0$
    \While{$iteration < \text{max\_iterations}$}
        \State Compute normalized values:
        \State $\hat{N}_m \gets \frac{{N_m} - {N_m}_{\text{min}}}{{N_m}_{\text{max}} - {N_m}_{\text{min}}}$
        \State $\hat{P}_e \gets \frac{{P_e} - {P_e}_{\text{min}}}{{P_e}_{\text{max}} - {P_e}_{\text{min}}}$
        
        \State Compute gradients:
        \State $d\_\hat{N}_m \gets \frac{w_{{N_m}}}{{N_m}_{\text{max}} - {N_m}_{\text{min}}}$
        \State $d\_\hat{P}_e \gets \frac{w_{{P_e}}}{{P_e}_{\text{max}} - {P_e}_{\text{min}}} \times \frac{{\partial {P_e}}}{{\partial {N_m}}}$ \Comment{from equation (\ref{partial_{P_e}})}
        
        \State Update parameters:
        \State ${N_m} \gets {N_m} - \text{learning\_rate} \times d\_\hat{N}_m$
        \State $P_e \gets P_e - \text{learning\_rate} \times d\_\hat{P}_e$
        
        \State Check for convergence:
        \If{$|f_{balance}({N_m}, {P_e}) - f_{balance}({N_m} - f_{balance} \times d\_\hat{N}_m, {P_e} - f_{balance} \times d\_\hat{P}_e)| < \text{tolerance}$}
            \State \textbf{break}
        \EndIf
        
        \State $iteration \gets iteration + 1$
    \EndWhile
    \State \textbf{return} ${N_m}, {P_e}$
\EndFunction
\end{algorithmic}
\end{algorithm}

\section{Numerical and Simulation Results}

In this section, we analyze the performance of the considered MC system, focusing on the trade-off between the number of transmitted molecules $N_m$ and the BER performance. This analysis is crucial as it helps in understanding how varying the number of transmitted molecules impacts the overall efficiency and reliability of the MC system. As the unit of SNR in the simulations is dB, the SNR is redefined as ${\rm{SNR}}=10\rm{log}\frac{y^2(t)}{\sigma_n^2}$. In the simulation, the parameters are set according to references \cite{musa2022optimized}, \cite{jing2024performance} and are presented in Table 1.
\begin{table}
\normalsize
\caption{SIMULATION PARAMETERS}
\centering
\begin{center}
\renewcommand\arraystretch{1.5} 
\setlength{\tabcolsep}{0.8mm}
\resizebox{0.5\textwidth}{!}{
\begin{tabular}{  c  c  c  p{3cm}}
\hline
Symbol & Explanation & Value  \\ \hline
$D_m$ & Diffusion coefficient & $10^{-9}$ m$^2$/s \\ \hline
$d$ & Distance between transmitter and receiver & 10 $\mu$m  \\ \hline
$r$ & Radius of the receiver & 4 $\mu$m \\ \hline
$T$ & Bit interval & 1 s \\ \hline
$\Delta t$ & Discrete steps & 100 $\mu s$ \\ \hline
$k$ & Boltzmann's constant & 1.3807 $\times$ $10^{-23}$ \\ \hline
$T_e$ & Absolute temperature & 298.15 \\ \hline
$L$ & Memory length & 4 \\ \hline
\end{tabular}}
\end{center}
\end{table}

\begin{figure}[!t]
   \centering
   \includegraphics[width=0.5\textwidth]{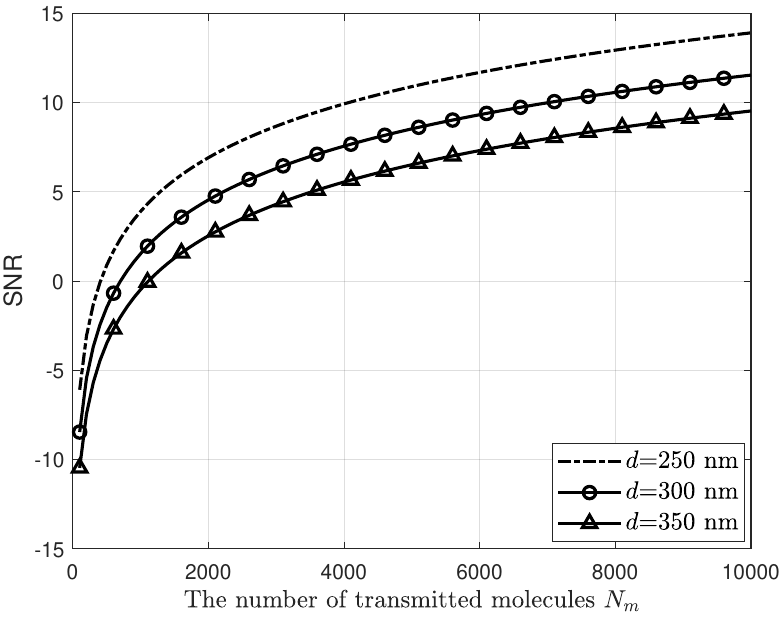}\\
   \caption{The SNR versus the number of transmitted molecules $N_m$ under different distances between the transmitter and receiver.}\label{SNR_Nm_d} 
\end{figure}

In Fig. \ref{SNR_Nm_d}, we analyze how the SNR varies with the number of transmitted molecules $N_m$ at different distances between the transmitter and the receiver. The SNR increases with the number of transmitted molecules, as indicated by (\ref{SNR}). This relationship arises because the variance of the noise is proportional to the number of transmitted molecules, a principle also applicable in electromagnetic communication. Consequently, higher numbers of transmitted molecules result in higher SNR values, similar to the relationship between SNR and transmission power. Additionally, larger distances between the transmitter and the receiver lead to a decrease in the peak of the channel impulse response, resulting in fewer received molecules. Therefore, the SNR is smaller at greater distances.

\begin{figure}[!t]
   \centering
   \includegraphics[width=0.5\textwidth]{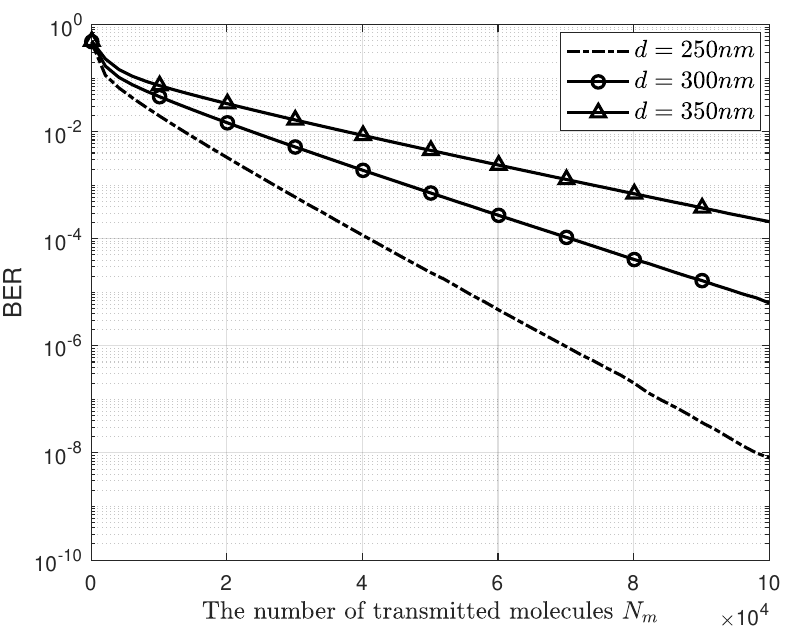}\\
   \caption{The BER versus the number of transmitted molecules $ N_m$ under different distances between the transmitter and receiver without considering the ISI.}\label{BER_Nm_d_without_ISI} 
\end{figure}

\begin{figure}[!t]
   \centering
   \includegraphics[width=0.5\textwidth]{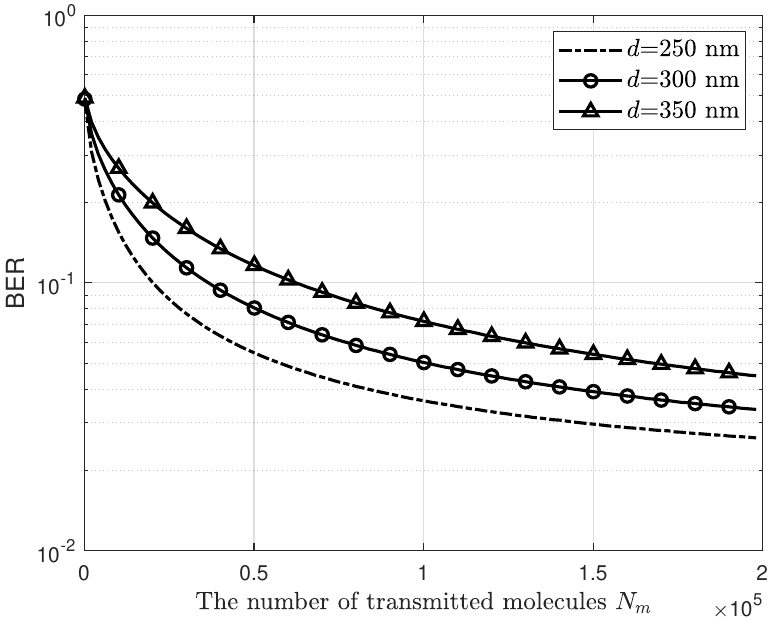}\\
   \caption{The BER versus the number of transmitted molecules $ N_m$ under different distances between the transmitter and receiver considering the ISI.}\label{BER_Nm_d} 
\end{figure}

In Fig. \ref{BER_Nm_d_without_ISI} and Fig. \ref{BER_Nm_d}, we illustrate the simulation results depicting the variation of the BER concerning the number of transmitted molecules $N_m$ across various distances $d$ between the transmitter and the receiver. While Fig. \ref{BER_Nm_d_without_ISI} excludes consideration of ISI, it is accounted for in Fig. \ref{BER_Nm_d}. The BER decreases with increasing transmitted molecules, corresponding to Fig. \ref{SNR_Nm_d} where the SNR improves with increased $N_m$, resulting in better BER performance. Additionally, for a given number of transmitted molecules, larger distances lead to lower SNR and consequently higher BER. Moreover, as shown in Fig. \ref{BER_Nm_d}, the rate of BER decrease slows down with increasing transmitted molecules, suggesting that blindly increasing the number of transmitted molecules does not necessarily improve BER performance. This highlights the importance of judicious use of transmitted molecules, a valuable resource in MC.

\begin{figure}[!t]
   \centering
   \includegraphics[width=0.5\textwidth]{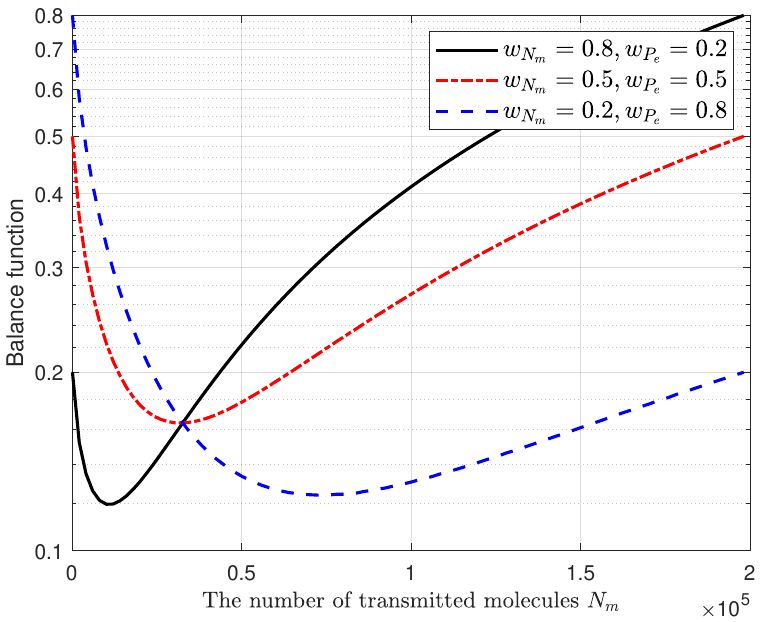}\\
   \caption{The balance function varies with the number of transmitted molecules $N_m$ under different weights of transmitted molecules and the BER performance.}\label{Balance_function} 
\end{figure}

\begin{figure}[!t]
   \centering
   \includegraphics[width=0.5\textwidth]{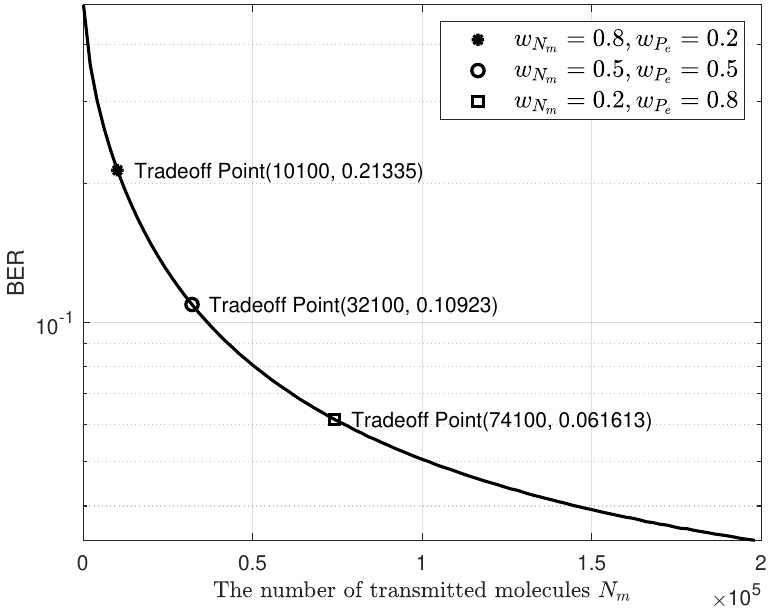}\\
   \caption{The tradeoff point under different weights of transmitted molecules and the BER performance.}\label{Tradeoff_point} 
\end{figure}

In Fig. \ref{Balance_function}, the defined balance function varies with the number of transmitted molecules under different weights of transmitted molecules \(N_m\), and BER \(P_e\) are simulated. As shown in the figure, for varying weights of \(N_m\) and \(P_e\), denoted as \(w_{N_m}\) and \(w_{P_e}\) respectively, it achieves different extreme points. This variation arises because the balance function assigns different emphases to its components for different weights. For example, when \(w_{N_m}\) is relatively large, the balance function prioritizes the number of molecules, leading to the attainment of the minimum point at smaller \(N_m\). Additionally, it is observed that the balance function initially decreases, then increases, before reaching the minimum point. In Fig. \ref{Tradeoff_point}, the tradeoff point under different \(w_{N_m}\) and \(w_{P_e}\) is simulated, representing the minimum point of the balance function. At this point, the BER is lower than the preset threshold while minimizing the number of transmitted molecules. As illustrated in Fig. \ref{Tradeoff_point}, the BER decreases as the number of transmitted molecules increases. For larger values of \(w_{N_m}\), the balance function prioritizes minimizing the number of molecules, resulting in fewer transmitted molecules but higher BER. Conversely, for larger values of \(w_{P_e}\), the balance function prioritizes minimizing the BER, leading to a greater number of transmitted molecules but lower BER. This result provides a basis for selecting an optimal tradeoff between the number of transmitted molecules and BER performance, depending on the specific requirements of the communication system.

\section{Conclusions}

{\color{black}In this paper, we investigated a diffusion-based MC system and analyzed how the BER varies with the number of transmitted molecules. Recognizing the significant difference in magnitude between the number of transmitted molecules and the BER, we initially normalized these two parameters. Given the scarcity of molecular resources in MC systems and the tendency for BER to decrease with increasing transmitted molecules, we proposed a balance function that jointly considers both factors. Using the gradient descent algorithm, we identified the minimum point, referred to as the tradeoff point, which offers a strategic choice between the number of transmitted molecules and BER performance. In MC systems with abundant molecular resources and a priority on achieving superior BER performance, releasing more molecules may be preferable. Conversely, in MC systems with limited molecular resources where achieving a very low BER is not critical, releasing a smaller number of molecules may suffice.} The BER of an MC system is indeed influenced by the nano receiver circuitry, which has not been factored into the current BER analysis. A more comprehensive BER definition and analysis will be explored in future work.

\appendix 
\section{}
As $Q\left( x \right) = \frac{1}{{\sqrt {2\pi } }}\int_x^\infty  {\exp \left( { - \frac{{{x^2}}}{2}} \right)} dx$, then
\begin{align}
\frac{{\partial Q}}{{\partial x}} = \frac{x}{{\sqrt {2\pi } }}\exp ( - \frac{{{x^2}}}{2}).
\end{align}
Based on (\ref{P_M}), $\frac{{\partial {P_M}}}{{\partial {N_m}}}$ can be expressed as
\begin{align}
\label{PM_NM}
\frac{{\partial {P_M}}}{{\partial {N_m}}} =  - \frac{1}{{{2^L}}}\sum\limits_{{b_L} \in \{ 0,1\} } {\Big[\frac{{\partial Q}}{{\partial (\frac{{{C_{\rm{Thr}}} - {\mu _{{Z_1}}}}}{{{\sigma _{{Z_1}}}}})}}}  \cdot \frac{{\partial (\frac{{{C_{\rm{Thr}}} - {\mu _{{Z_1}}}}}{{{\sigma _{{Z_1}}}}})}}{{\partial {N_m}}}\Big],
\end{align}
where 
\begin{align}
{\mu _{{Z_1}}} = {N_m}{h_1} + \sum\limits_{i = k - L}^k {{b_i}{N_m}h((k - i){T_b} + {t_p})},
\end{align}
and
\begin{align}
{\sigma _{{Z_1}}} = \sqrt {\frac{{{b_1}{N_m}{h_1}}}{{{V_R}}} + \frac{1}{{{V_R}}}\sum\limits_{i = k - L}^k {{b_i}{N_m}h((k - i){T_b} + {t_p})} }.
\end{align}
Then, $\frac{{\partial {\mu _{{Z_1}}}}}{{\partial {N_m}}}$ and $\frac{{\partial {\sigma _{{Z_1}}}}}{{\partial {N_m}}}$ can be expressed as
\begin{align}
\frac{{\partial {\mu _{{Z_1}}}}}{{\partial {N_m}}} = {h_1} + \sum\limits_{i = k - L}^k {{b_i}h((k - i){T_b} + {t_p})},
\end{align}
and
\begin{align}
\frac{{\partial {\sigma _{{Z_1}}}}}{{\partial {N_m}}} = \frac{1}{{2{\sigma _{{Z_1}}}}} \cdot (\frac{{{b_1}{h_1}}}{{{V_R}}} + \frac{1}{{{V_R}}}\sum\limits_{i = k - L}^k {{b_i}h((k - i){T_b} + {t_p})} ),
\end{align}
respectively. Therefore, in (\ref{PM_NM}), $\frac{{\partial Q}}{{\partial (\frac{{{C_{\rm{Thr}}} - {\mu _{{Z_1}}}}}{{{\sigma _{{Z_1}}}}})}}$ and $\frac{{\partial (\frac{{{C_{\rm{Thr}}} - {\mu _{{Z_1}}}}}{{{\sigma _{{Z_1}}}}})}}{{\partial {N_m}}}$ can be expressed as
\begin{align}
\frac{{\partial Q}}{{\partial (\frac{{{C_{\rm{Thr}}} - {\mu _{{Z_1}}}}}{{{\sigma _{{Z_1}}}}})}} = \frac{{{C_{\rm{Thr}}} - {\mu _{{Z_1}}}}}{{{\sigma _{{Z_1}}}\sqrt {2\pi } }} \cdot \exp ( - {{{{(\frac{{{C_{\rm{Thr}}} - {\mu _{{Z_1}}}}}{{{\sigma _{{Z_1}}}}})}^2}} \mathord{\left/
 {\vphantom {{{{(\frac{{{C_{\rm{Thr}}} - {\mu _{{Z_1}}}}}{{{\sigma _{{Z_1}}}}})}^2}} 2}} \right.
 \kern-\nulldelimiterspace} 2}),
\end{align}
and 
\begin{align}
\frac{{\partial (\frac{{{C_{\rm{Thr}}} - {\mu _{{Z_1}}}}}{{{\sigma _{{Z_1}}}}})}}{{\partial {N_m}}} = \frac{{ - \frac{{\partial {\mu _{{Z_1}}}}}{{\partial {N_m}}}{\sigma _{{Z_1}}} - ({C_{\rm{Thr}}} - {\mu _{{Z_1}}})\frac{{\partial {\sigma _{{Z_1}}}}}{{\partial {N_m}}}}}{{\sigma _{{Z_1}}^2}},
\end{align}
respectively. Then, $\frac{{\partial {P_M}}}{{\partial {N_m}}}$ can be expressed as
\begin{align}
\begin{split}
&\frac{{\partial {P_M}}}{{\partial {N_m}}} =  - \frac{1}{{{2^L}}}\sum\limits_{{b_L} \in \{ 0,1\} } {\Big[\frac{{\partial Q}}{{\partial (\frac{{{C_{\rm{Thr}}} - {\mu _{{Z_1}}}}}{{{\sigma _{{Z_1}}}}})}}}  \cdot \frac{{\partial (\frac{{{C_{\rm{Thr}}} - {\mu _{{Z_1}}}}}{{{\sigma _{{Z_1}}}}})}}{{\partial {N_m}}}\Big]\\
&= - \frac{1}{{{2^L}}}\sum\limits_{{b_L} \in \{ 0,1\} } {\Big[\Big(\frac{{{C_{\rm{Thr}}} - {\mu _{{Z_1}}}}}{{{\sigma _{{Z_1}}}\sqrt {2\pi } }} \exp \Big( - {{{{(\frac{{{C_{\rm{Thr}}} - {\mu _{{Z_1}}}}}{{{\sigma _{{Z_1}}}}})}^2}} \mathord{\left/
 {\vphantom {{{{(\frac{{{C_{\rm{Thr}}} - {\mu _{{Z_1}}}}}{{{\sigma _{{Z_1}}}}})}^2}} 2}} \right.
 \kern-\nulldelimiterspace} 2}\Big)\Big)}\\
 &\times \frac{{ - \frac{{\partial {\mu _{{Z_1}}}}}{{\partial {N_m}}}{\sigma _{{Z_1}}} - ({C_{\rm{Thr}}} - {\mu _{{Z_1}}}) \frac{{\partial {\sigma _{{Z_1}}}}}{{\partial {N_m}}}}}{{\sigma _{{Z_1}}^2}}\Big]\\
& = \frac{1}{{{2^L}}}\sum\limits_{{b_L} \in \{ 0,1\} } {\Big[\Big(\frac{{{C_{\rm{Thr}}} - {\mu _{{Z_1}}}}}{{{\sigma _{{Z_1}}}\sqrt {2\pi } }} \exp \Big( - {{{{(\frac{{{C_{\rm{Thr}}} - {\mu _{{Z_1}}}}}{{{\sigma _{{Z_1}}}}})}^2}} \mathord{\left/
 {\vphantom {{{{(\frac{{{C_{Thr}} - {\mu _{{Z_1}}}}}{{{\sigma _{{Z_1}}}}})}^2}} 2}} \right.
 \kern-\nulldelimiterspace} 2}\Big)\Big)}\\
 &\times \frac{{\frac{{\partial {\mu _{{Z_1}}}}}{{\partial {N_m}}}{\sigma _{{Z_1}}} + ({C_{\rm{Thr}}} - {\mu _{{Z_1}}})\frac{{\partial {\sigma _{{Z_1}}}}}{{\partial {N_m}}}}}{{\sigma _{{Z_1}}^2}}\Big].
\end{split}
\end{align}
Similarly, Based on (\ref{P_FA}), $\frac{{\partial {P_{FA}}}}{{\partial {N_m}}}$ can be expressed as
\begin{align}
\label{FA_NM}
\frac{{\partial {P_{FA}}}}{{\partial {N_m}}} = \frac{1}{{{2^L}}}\sum\limits_{{b_L} \in \{ 0,1\} } {\Big[\frac{{\partial Q}}{{\partial (\frac{{{C_{\rm{Thr}}} - {\mu _{{Z_0}}}}}{{{\sigma _{{Z_0}}}}})}}}  \cdot \frac{{\partial (\frac{{{C_{\rm{Thr}}} - {\mu _{{Z_0}}}}}{{{\sigma _{{Z_0}}}}})}}{{\partial {N_m}}}\Big],
\end{align}
where
\begin{align}
{\mu _{{Z_0}}} = \sum\limits_{i = k - L}^k {{b_i}{N_m}h((k - i){T_b} + {t_p})}, 
\end{align}
and
\begin{align}
{\sigma _{{Z_0}}} = \sqrt {\frac{1}{{{V_R}}}\sum\limits_{i = k - L}^k {{b_i}{N_m}h((k - i){T_b} + {t_p})} } .
\end{align}
Then, $\frac{{\partial {\mu _{{Z_0}}}}}{{\partial {N_m}}}$ and $\frac{{\partial {\sigma _{{Z_0}}}}}{{\partial {N_m}}}$ can be expressed as
\begin{align}
\frac{{\partial {\mu _{{Z_0}}}}}{{\partial {N_m}}} = \sum\limits_{i = k - L}^k {{b_i}h((k - i){T_b} + {t_p})} ,
\end{align}

\begin{align}
\frac{{\partial {\sigma _{{Z_0}}}}}{{\partial {N_m}}} = \frac{1}{{2{\sigma _{{Z_0}}}}} \cdot (\frac{1}{{{V_R}}}\sum\limits_{i = k - L}^k {{b_i}h((k - i){T_b} + {t_p})} ),
\end{align}
respectively. Therefore, in (\ref{FA_NM}), $\frac{{\partial (\frac{{{C_{\rm{Thr}}} - {\mu _{{Z_0}}}}}{{{\sigma _{{Z_0}}}}})}}{{\partial {N_m}}}$ and $\frac{{\partial Q}}{{\partial (\frac{{{C_{\rm{Thr}}} - {\mu _{{Z_0}}}}}{{{\sigma _{{Z_0}}}}})}}$ can be expressed as
\begin{align}
\frac{{\partial (\frac{{{C_{\rm{Thr}}} - {\mu _{{Z_0}}}}}{{{\sigma _{{Z_0}}}}})}}{{\partial {N_m}}} = \frac{{ - \frac{{\partial {\mu _{{Z_0}}}}}{{\partial {N_m}}}{\sigma _{{Z_0}}} - ({C_{Thr}} - {\mu _{{Z_0}}}) \frac{{\partial {\sigma _{{Z_0}}}}}{{\partial {N_m}}}}}{{\sigma _{{Z_0}}^2}},
\end{align}

\begin{align}
\begin{split}
\frac{{\partial Q}}{{\partial (\frac{{{C_{\rm{Thr}}} - {\mu _{{Z_0}}}}}{{{\sigma _{{Z_0}}}}})}} = \frac{{{C_{\rm{Thr}}} - {\mu _{{Z_0}}}}}{{{\sigma _{{Z_0}}}\sqrt {2\pi } }}\exp \Big( - {{{{(\frac{{{C_{\rm{Thr}}} - {\mu _{{Z_0}}}}}{{{\sigma _{{Z_0}}}}})}^2}} \mathord{\left/
 {\vphantom {{{{(\frac{{{C_{\rm{Thr}}} - {\mu _{{Z_0}}}}}{{{\sigma _{{Z_0}}}}})}^2}} 2}} \right.
 \kern-\nulldelimiterspace} 2}\Big),
\end{split}
\end{align}
respectively. Then $\frac{{\partial {P_{FA}}}}{{\partial {N_m}}}$ can be expressed as
\begin{align}
\begin{split}
&\frac{{\partial {P_{FA}}}}{{\partial {N_m}}} = \frac{1}{{{2^L}}}\sum\limits_{{b_L} \in \{ 0,1\} } {\Big[\frac{{\partial Q}}{{\partial (\frac{{{C_{\rm{Thr}}} - {\mu _{{Z_0}}}}}{{{\sigma _{{Z_0}}}}})}}}  \cdot \frac{{\partial (\frac{{{C_{\rm{Thr}}} - {\mu _{{Z_0}}}}}{{{\sigma _{{Z_0}}}}})}}{{\partial {N_m}}}\Big]\\
&= \frac{1}{{{2^L}}}\sum\limits_{{b_L} \in \{ 0,1\} } {\Big[\Big(\frac{{{C_{\rm{Thr}}} - {\mu _{{Z_0}}}}}{{{\sigma _{{Z_0}}}\sqrt {2\pi } }}\exp \Big( - {{{{(\frac{{{C_{Thr}} - {\mu _{{Z_0}}}}}{{{\sigma _{{Z_0}}}}})}^2}} \mathord{\left/
 {\vphantom {{{{(\frac{{{C_{\rm{Thr}}} - {\mu _{{Z_0}}}}}{{{\sigma _{{Z_0}}}}})}^2}} 2}} \right.
 \kern-\nulldelimiterspace} 2}\Big)\Big)}\\
 &\times \frac{{ - {\kern 1pt} \frac{{\partial {\mu _{{Z_0}}}}}{{\partial {N_m}}}{\sigma _{{Z_0}}} - ({C_{\rm{Thr}}} - {\mu _{{Z_0}}}){\kern 1pt} \frac{{\partial {\sigma _{{Z_0}}}}}{{\partial {N_m}}}}}{{\sigma _{{Z_0}}^2}}\Big]\\
& = - \frac{1}{{{2^L}}}\sum\limits_{{b_L} \in \{ 0,1\} } {\Big[\Big(\frac{{{C_{\rm{Thr}}} - {\mu _{{Z_0}}}}}{{{\sigma _{{Z_0}}}\sqrt {2\pi } }}\exp \Big( - {{{{(\frac{{{C_{\rm{Thr}}} - {\mu _{{Z_0}}}}}{{{\sigma _{{Z_0}}}}})}^2}} \mathord{\left/
 {\vphantom {{{{(\frac{{{C_{Thr}} - {\mu _{{Z_0}}}}}{{{\sigma _{{Z_0}}}}})}^2}} 2}} \right.
 \kern-\nulldelimiterspace} 2}\Big)\Big)}  \\
&\times \frac{{\frac{{\partial {\mu _{{Z_0}}}}}{{\partial {N_m}}}{\sigma _{{Z_0}}} + ({C_{\rm{Thr}}} - {\mu _{{Z_0}}}){\kern 1pt} \frac{{\partial {\sigma _{{Z_0}}}}}{{\partial {N_m}}}}}{{\sigma _{{Z_0}}^2}}\Big]
\end{split}
\end{align}

{\color{black}\bibliographystyle{IEEEtran}
\bibliography{references}}

\begin{thebibliography}{10}
\providecommand{\url}[1]{#1}
\csname url@samestyle\endcsname
\providecommand{\newblock}{\relax}
\providecommand{\bibinfo}[2]{#2}
\providecommand{\BIBentrySTDinterwordspacing}{\spaceskip=0pt\relax}
\providecommand{\BIBentryALTinterwordstretchfactor}{4}
\providecommand{\BIBentryALTinterwordspacing}{\spaceskip=\fontdimen2\font plus
\BIBentryALTinterwordstretchfactor\fontdimen3\font minus
  \fontdimen4\font\relax}
\providecommand{\BIBforeignlanguage}[2]{{%
\expandafter\ifx\csname l@#1\endcsname\relax
\typeout{** WARNING: IEEEtran.bst: No hyphenation pattern has been}%
\typeout{** loaded for the language `#1'. Using the pattern for}%
\typeout{** the default language instead.}%
\else
\language=\csname l@#1\endcsname
\fi
#2}}
\providecommand{\BIBdecl}{\relax}
\BIBdecl

\bibitem{farsad2016comprehensive}
N.~Farsad, H.~B. Yilmaz, A.~Eckford, C.-B. Chae, and W.~Guo, ``A comprehensive
  survey of recent advancements in molecular communication,'' \emph{IEEE
  Communications Surveys \& Tutorials}, vol.~18, no.~3, pp. 1887--1919, 2016.

\bibitem{saeed2023novel}
M.~Saeed, M.~Maleki, and H.~R. Bahrami, ``A novel inter-symbol interference
  model and weighted sum detection for diffusion-based molecular communication
  systems,'' \emph{IEEE Transactions on Molecular, Biological and Multi-Scale
  Communications}, 2023.

\bibitem{jing2024energy}
D.~Jing, L.~Lin, and A.~W. Eckford, ``Energy allocation for multi-user
  cooperative molecular communication systems in the internet of bio-nano
  things,'' \emph{IEEE Internet of Things Journal}, 2024.

\bibitem{10443866}
M.~Hamidović, S.~Angerbauer, D.~Bi, Y.~Deng, T.~Tugcu, and W.~Haselmayr,
  ``Microfluidic systems for molecular communications: A review from theory to
  practice,'' \emph{IEEE Transactions on Molecular, Biological, and Multi-Scale
  Communications}, vol.~10, no.~1, pp. 147--163, 2024.

\bibitem{yang2020comprehensive}
K.~Yang, D.~Bi, Y.~Deng, R.~Zhang, M.~M.~U. Rahman, N.~A. Ali, M.~A. Imran,
  J.~M. Jornet, Q.~H. Abbasi, and A.~Alomainy, ``A comprehensive survey on
  hybrid communication in context of molecular communication and terahertz
  communication for body-centric nanonetworks,'' \emph{IEEE Transactions on
  Molecular, Biological and Multi-Scale Communications}, vol.~6, no.~2, pp.
  107--133, 2020.

\bibitem{khaloopour2021theoretical}
L.~Khaloopour, M.~Mirmohseni, and M.~Nasiri-Kenari, ``Theoretical concept study
  of cooperative abnormality detection and localization in fluidic-medium
  molecular communication,'' \emph{IEEE sensors journal}, vol.~21, no.~15, pp.
  17\,118--17\,130, 2021.

\bibitem{zhang2024design}
C.~Zhang, H.~Yan, Q.~Liu, K.~Yang, F.~Liu, and L.~Lin, ``Design and analysis of
  a through-body signal transmission system based on human oxygen saturation
  detection,'' \emph{IEEE Transactions on Molecular, Biological and Multi-Scale
  Communications}, 2024.

\bibitem{xiao2023really}
H.~Xiao, K.~Dokaj, and O.~B. Akan, ``What really is ‘molecule’in molecular
  communications? the quest for physics of particle-based information
  carriers,'' \emph{IEEE Transactions on Molecular, Biological and Multi-Scale
  Communications}, 2023.

\bibitem{kuran2011modulation}
M.~S. Kuran, H.~B. Yilmaz, T.~Tugcu, and I.~F. Akyildiz, ``Modulation
  techniques for communication via diffusion in nanonetworks,'' in \emph{2011
  IEEE international conference on communications (ICC)}.\hskip 1em plus 0.5em
  minus 0.4em\relax IEEE, 2011, pp. 1--5.

\bibitem{4298292}
A.~W. Eckford, ``Nanoscale communication with brownian motion,'' in \emph{2007
  41st Annual Conference on Information Sciences and Systems}, 2007, pp.
  160--165.

\bibitem{li2019asymmetric}
Q.~Li, ``The asymmetric-distance metrics for decoding of convolutional codes in
  diffusion-based molecular communications,'' \emph{IEEE Transactions on
  NanoBioscience}, vol.~18, no.~3, pp. 469--481, 2019.

\bibitem{nakano2012molecular}
T.~Nakano, M.~J. Moore, F.~Wei, A.~V. Vasilakos, and J.~Shuai, ``Molecular
  communication and networking: Opportunities and challenges,'' \emph{IEEE
  transactions on nanobioscience}, vol.~11, no.~2, pp. 135--148, 2012.

\bibitem{nakano2017molecular}
T.~Nakano and T.~Suda, ``Molecular communication using dynamic properties of
  oscillating and propagating patterns in concentration of information
  molecules,'' \emph{IEEE Transactions on Communications}, vol.~65, no.~8, pp.
  3386--3398, 2017.

\bibitem{moore2014diffusion}
M.~J. Moore, Y.~Okaie, and T.~Nakano, ``Diffusion-based multiple access by
  nano-transmitters to a micro-receiver,'' \emph{IEEE Communications Letters},
  vol.~18, no.~3, pp. 385--388, 2014.

\bibitem{li2019clock}
Q.~Li, ``The clock-free asynchronous receiver design for molecular timing
  channels in diffusion-based molecular communications,'' \emph{IEEE
  Transactions on NanoBioscience}, vol.~18, no.~4, pp. 585--596, 2019.

\bibitem{bi2021survey}
D.~Bi, A.~Almpanis, A.~Noel, Y.~Deng, and R.~Schober, ``A survey of molecular
  communication in cell biology: Establishing a new hierarchy for
  interdisciplinary applications,'' \emph{IEEE Communications Surveys \&
  Tutorials}, vol.~23, no.~3, pp. 1494--1545, 2021.

\bibitem{deng2016enabling}
Y.~Deng, W.~Guo, A.~Noel, A.~Nallanathan, and M.~Elkashlan, ``Enabling energy
  efficient molecular communication via molecule energy transfer,'' \emph{IEEE
  Communications Letters}, vol.~21, no.~2, pp. 254--257, 2016.

\bibitem{musa2020feedback}
V.~Musa, G.~Piro, L.~A. Grieco, and G.~Boggia, ``A feedback control strategy
  for energy-harvesting in diffusion-based molecular communication systems,''
  \emph{IEEE Transactions on Communications}, vol.~69, no.~2, pp. 831--844,
  2020.

\bibitem{lotter2020synaptic}
S.~Lotter, A.~Ahmadzadeh, and R.~Schober, ``Synaptic channel modeling for dmc:
  Neurotransmitter uptake and spillover in the tripartite synapse,'' \emph{IEEE
  Transactions on Communications}, vol.~69, no.~3, pp. 1462--1479, 2020.

\bibitem{ahmadzadeh2022molecule}
A.~Ahmadzadeh, V.~Jamali, and R.~Schober, ``Molecule harvesting transmitter
  model for molecular communication systems,'' \emph{IEEE Open Journal of the
  Communications Society}, vol.~3, pp. 391--410, 2022.

\bibitem{huang2023analysis}
X.~Huang, Y.~Huang, M.~Wen, N.~Yang, and R.~Schober, ``Analysis of molecule
  harvesting by heterogeneous receptors on mc transmitters,'' in \emph{2023
  IEEE Globecom Workshops (GC Wkshps)}.\hskip 1em plus 0.5em minus 0.4em\relax
  IEEE, 2023, pp. 545--551.

\bibitem{wen2024absorption}
M.~Wen, F.~Liang, W.~Ye, and X.~Chen, ``Absorption shift keying for molecular
  communication via diffusion,'' \emph{IEEE Transactions on Molecular,
  Biological and Multi-Scale Communications}, 2024.

\bibitem{kumar2024game}
S.~Kumar, P.~K. Sharma, and M.~R. Bhatnagar, ``Game-theoretic performance of
  molecular reporting channel with receiver nano-machine,'' \emph{IEEE
  Transactions on Molecular, Biological, and Multi-Scale Communications}, 2024.

\bibitem{nakano2013transmission}
T.~Nakano, Y.~Okaie, and A.~V. Vasilakos, ``Transmission rate control for
  molecular communication among biological nanomachines,'' \emph{IEEE Journal
  on Selected Areas in Communications}, vol.~31, no.~12, pp. 835--846, 2013.

\bibitem{khaloopour2018adaptive}
L.~Khaloopour, M.~Mirmohseni, and M.~Nasiri-Kenari, ``An adaptive pulse-width
  modulation for limited molecule production and storage,'' in \emph{2018 Iran
  Workshop on Communication and Information Theory (IWCIT)}.\hskip 1em plus
  0.5em minus 0.4em\relax IEEE, 2018, pp. 1--6.

\bibitem{cheng2023channel}
Z.~Cheng, J.~Sun, Z.~Zhang, P.~Hu, and K.~Chi, ``Channel modeling and optimal
  released molecules for mobile molecular mimo communications among
  bionanosensors,'' \emph{IEEE Sensors Journal}, 2023.

\bibitem{panahi2024energy}
F.~H. Panahi and F.~H. Panahi, ``Energy-efficient data collection in molecular
  nanonetworks: An optimization framework,'' \emph{IEEE Signal Processing
  Letters}, 2024.

\bibitem{cheng2022joint}
Z.~Cheng, J.~Yan, J.~Sun, Y.~Tu, and K.~Chi, ``Joint optimizations of relays
  locations and decision threshold for multi-hop diffusive mobile molecular
  communication with drift,'' \emph{IEEE Transactions on NanoBioscience},
  vol.~21, no.~3, pp. 454--465, 2022.

\bibitem{shitiri2021tdma}
E.~Shitiri and H.-S. Cho, ``A tdma-based data gathering protocol for molecular
  communication via diffusion-based nano-sensor networks,'' \emph{IEEE Sensors
  Journal}, vol.~21, no.~17, pp. 19\,582--19\,595, 2021.

\bibitem{li2019novel}
Q.~Li, ``A novel time-based modulation scheme in time-asynchronous channels for
  molecular communications,'' \emph{IEEE Transactions on NanoBioscience},
  vol.~19, no.~1, pp. 59--67, 2019.

\bibitem{chouhan2020gradient}
L.~Chouhan, N.~Varshney, and P.~K. Sharma, ``On gradient descent optimization
  in diffusion-advection based 3-d molecular cooperative communication,''
  \emph{IEEE Transactions on NanoBioscience}, vol.~19, no.~3, pp. 347--356,
  2020.

\bibitem{pierobon2011diffusion}
M.~Pierobon and I.~F. Akyildiz, ``Diffusion-based noise analysis for molecular
  communication in nanonetworks,'' \emph{IEEE Transactions on signal
  processing}, vol.~59, no.~6, pp. 2532--2547, 2011.

\bibitem{kuscu2019transmitter}
M.~Kuscu, E.~Dinc, B.~A. Bilgin, H.~Ramezani, and O.~B. Akan, ``Transmitter and
  receiver architectures for molecular communications: A survey on physical
  design with modulation, coding, and detection techniques,'' \emph{Proceedings
  of the IEEE}, vol. 107, no.~7, pp. 1302--1341, 2019.

\bibitem{trinh2021molecular}
D.~P. Trinh, Y.~Jeong, and S.-H. Kim, ``Molecular communication with passive
  receivers in anomalous diffusion channels,'' \emph{IEEE Wireless
  Communications Letters}, vol.~10, no.~10, pp. 2215--2219, 2021.

\bibitem{wang2015distance}
X.~Wang, M.~D. Higgins, and M.~S. Leeson, ``Distance estimation schemes for
  diffusion based molecular communication systems,'' \emph{IEEE Communications
  Letters}, vol.~19, no.~3, pp. 399--402, 2015.

\bibitem{sabu2019analysis}
N.~V. Sabu and A.~K. Gupta, ``Analysis of diffusion based molecular
  communication with multiple transmitters having individual random information
  bits,'' \emph{IEEE Transactions on Molecular, Biological and Multi-Scale
  Communications}, vol.~5, no.~3, pp. 176--188, 2019.

\bibitem{8412141}
L.~Shi and L.-L. Yang, ``Error performance analysis of diffusive molecular
  communication systems with on-off keying modulation,'' \emph{IEEE
  Transactions on Molecular, Biological and Multi-Scale Communications},
  vol.~3, no.~4, pp. 224--238, 2017.

\bibitem{liu2019efficient}
Y.~Liu, W.~Huangfu, H.~Zhang, and K.~Long, ``An efficient stochastic gradient
  descent algorithm to maximize the coverage of cellular networks,'' \emph{IEEE
  transactions on wireless communications}, vol.~18, no.~7, pp. 3424--3436,
  2019.

\bibitem{chen2021robust}
J.~Chen, M.~Gan, Q.~Zhu, P.~Narayan, and Y.~Liu, ``Robust standard gradient
  descent algorithm for arx models using aitken acceleration technique,''
  \emph{IEEE transactions on cybernetics}, vol.~52, no.~9, pp. 9646--9655,
  2021.

\bibitem{musa2022optimized}
V.~Musa, G.~Piro, L.~Grieco, and G.~Boggia, ``An optimized energy-harvesting
  transmission scheme for diffusion-based molecular communications,''
  \emph{IEEE Transactions on NanoBioscience}, vol.~22, no.~2, pp. 345--355,
  2022.

\bibitem{jing2024performance}
D.~Jing, L.~Lin, and A.~W. Eckford, ``Performance analysis and isi mitigation
  with imperfect transmitter in molecular communication,'' \emph{IEEE
  Transactions on NanoBioscience}, 2024.

\end{thebibliography}

\begin{IEEEbiography}[{\includegraphics[width=1in,height=1.25in,clip,keepaspectratio]{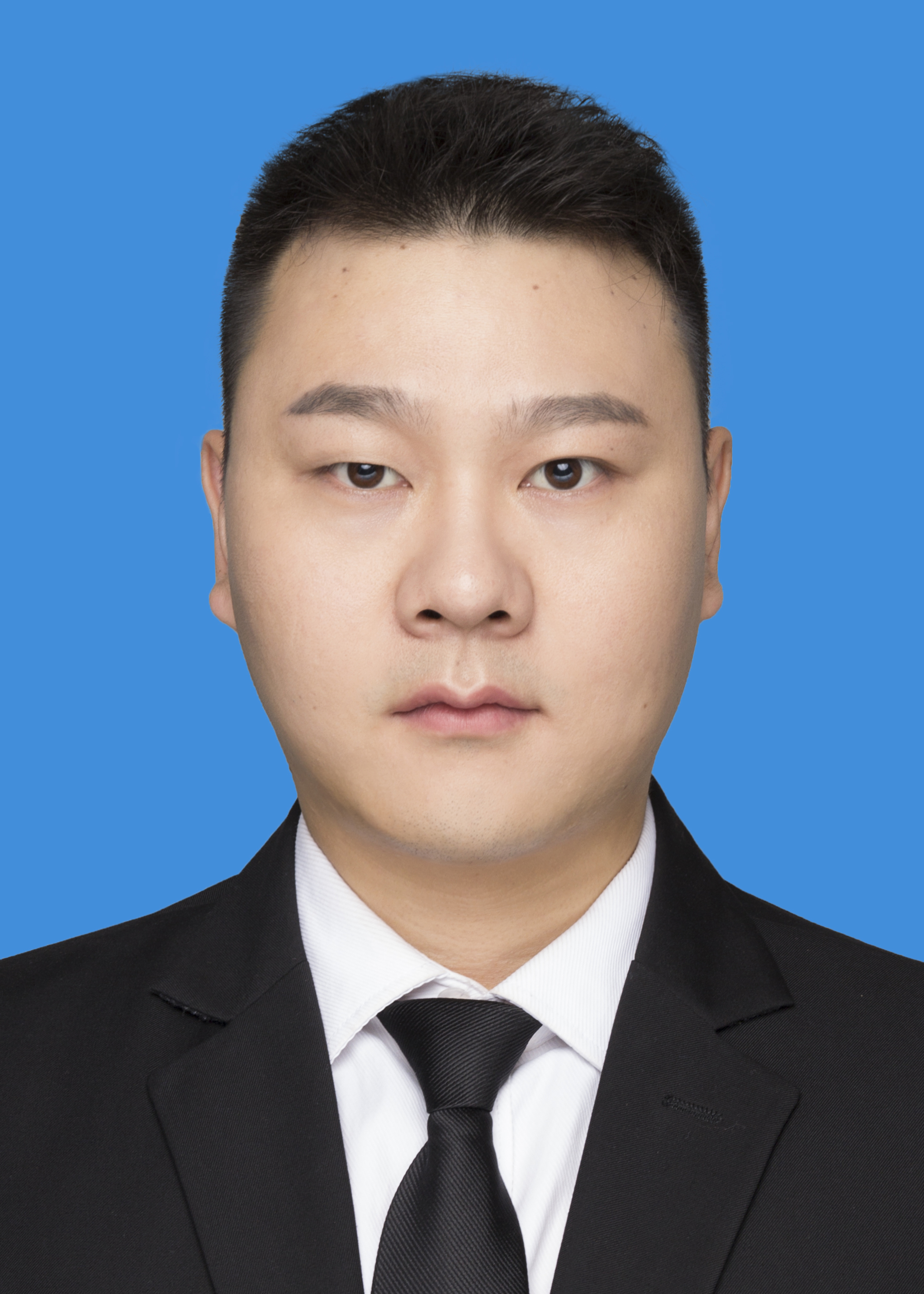}}]{Dongliang Jing} is a lecturer in the College of Mechanical and Electronic Engineering, Northwest A\&F University, Yangling, China. He received the B.S. degree from Anhui Polytechnic University, Wuhu, China in 2015 and received the Ph.D. degree from the Xidian University, Xian, China in 2022. During Nov. 2019 - Nov. 2020, he was a visiting student for molecular communication in York University, Toronto, ON. Canada, under the supervisor of Andrew W. Eckford. His main research interests include wireless and molecular communications.
\end{IEEEbiography}

\begin{IEEEbiography}[{\includegraphics[width=1in,height=1.25in,clip,keepaspectratio]{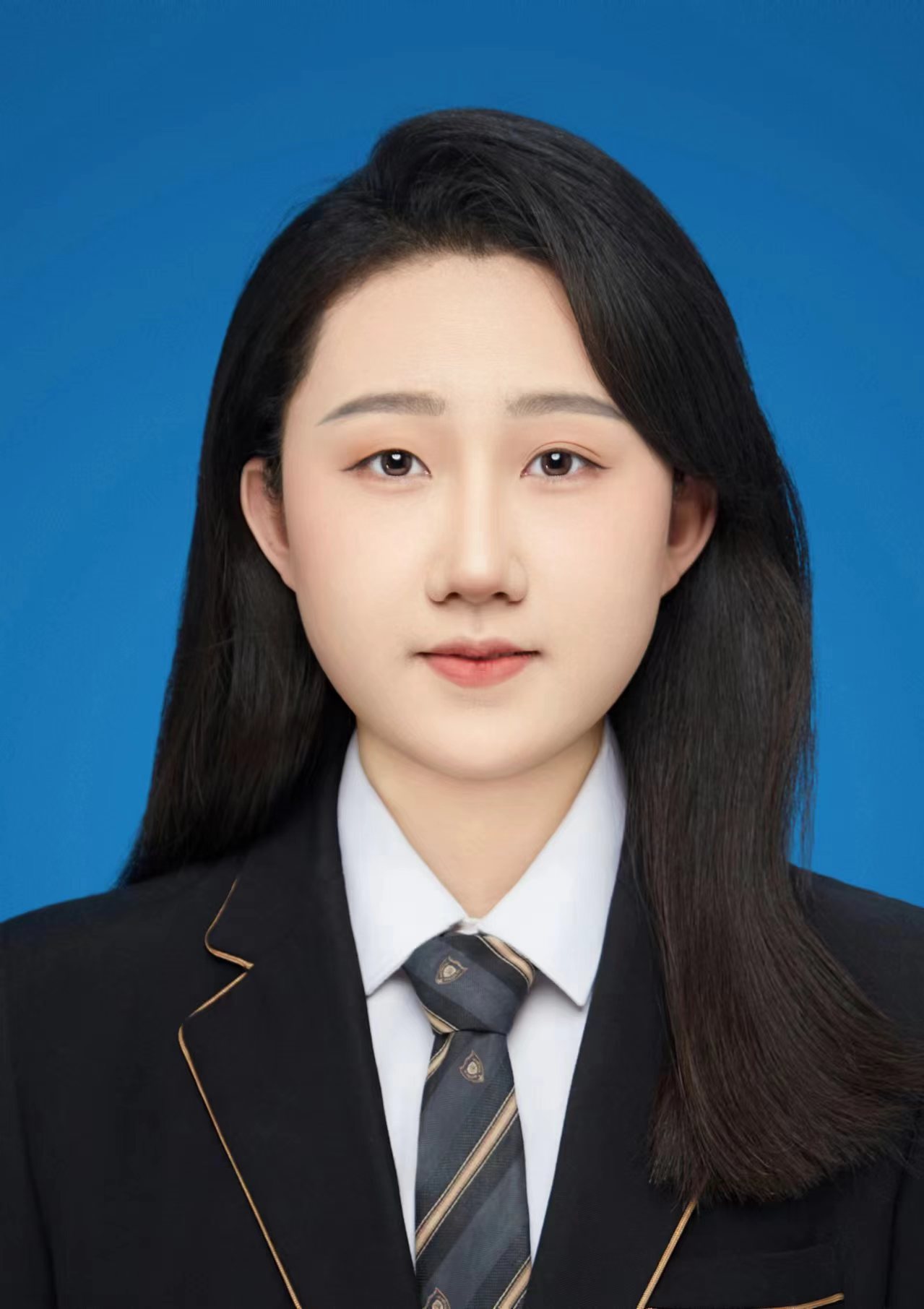}}]{Linjuan Li} is an undergraduate student in the College of Mechanical and Electronic Engineering, Northwest A\&F University. Her research interests include wireless and molecular communications.
\end{IEEEbiography}

\begin{IEEEbiography}[{\includegraphics[width=1in,height=1.25in, clip,keepaspectratio]{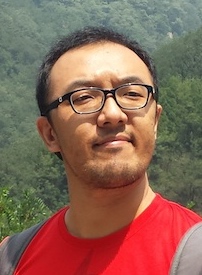}}]{Lin Lin} (Senior Member, IEEE) received the B.Eng. and M.Eng. degrees in electrical engineering from Tianjin University, China in 2004 and 2007, respectively, and the Ph.D. degree from Nanyang Technological University, Singapore in 2012. He is an Associate Professor with the College of Electronic and Information Engineering, Tongji University, Shanghai, China. He is currently the Chair of IEEE Nanotechnology Council Nanoscale Communications Technical Committee, and served as the Chair of IEEE ComSoc Molecular, Biological and Multi-scale Communications Technical Committee for the term 2022-2023. He serves as Associate Editor for IEEE Transactions on Molecular, Biological and Multi-scale Communications, IEEE Nanotechnology Magazine, and IEEE Access. His research interests include molecular communications, neural communications, internet of nanothings, and body sensor networks.
\end{IEEEbiography}

\begin{IEEEbiography}[{\includegraphics[width=1in,height=1.25in,clip,keepaspectratio]{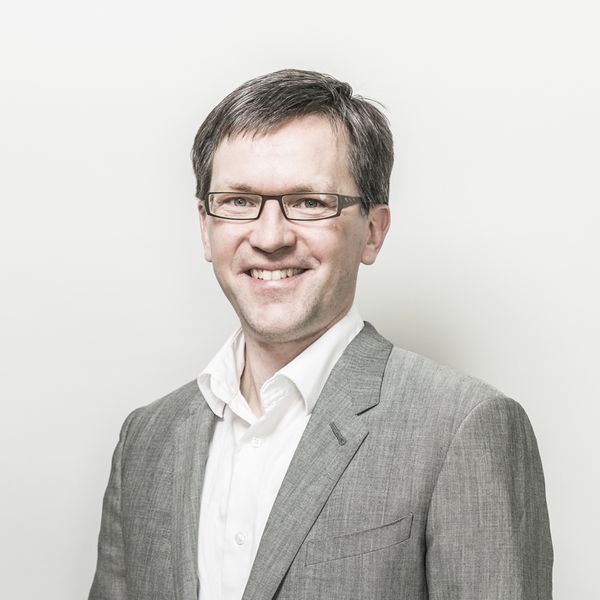}}]{Andrew W. Eckford} is an Associate Professor in the Department of Electrical Engineering and Computer Science at York University, Toronto, Ontario. His research interests include the application of information theory to biology, and the design of communication systems using molecular and biological techniques. His research has been covered in media including The Economist, The Wall Street Journal, and IEEE Spectrum. His research received the 2015 IET Communications Innovation Award, and was a finalist for the 2014 Bell Labs Prize. He is also a co-author of the textbook Molecular Communication, published by Cambridge University Press. Andrew received the B.Eng. degree from the Royal Military College of Canada in 1996, and the M.A.Sc. and Ph.D. degrees from the University of Toronto in 1999 and 2004, respectively, all in Electrical Engineering. Andrew held postdoctoral fellowships at the University of Notre Dame and the University of Toronto, prior to taking up a faculty position at York in 2006. He has held courtesy appointments at the University of Toronto and Case Western Reserve University. In 2018, he was named a Senior Fellow of Massey College, Toronto.
\end{IEEEbiography}
\end{document}